\documentclass[doublecol]{epl2}
\usepackage{color}
\usepackage{amsmath,amssymb}
\usepackage[dvipdfm,colorlinks,linkcolor=blue,citecolor=blue,urlcolor=blue,anchorcolor=blue]{hyperref}

\title{Fresnel diffraction patterns as accelerating beams}

\author{Yiqi Zhang\inst{1,(a)} \and Milivoj R. Beli\'c\inst{2} \and Huaibin Zheng\inst{1}
\and Zhenkun Wu\inst{1} \and Yuanyuan Li\inst{3} \and Keqing Lu\inst{4} \and Yanpeng Zhang\inst{1,(b)}}
\shortauthor{Y. Q. Zhang, M. R. Beli\'c, H. B. Zheng, Z. K. Wu, Y. Y. Li, K. Q. Lu, and Y. P. Zhang.}

\institute{
  \inst{1} Key Laboratory for Physical Electronics and Devices of the Ministry of Education \&
Shaanxi Key Lab of Information Photonic Technique,
Xi'an Jiaotong University, Xi'an 710049, China\\
  \inst{2} Science Program, Texas A\&M University at Qatar, P.O. Box 23874 Doha, Qatar \\
  \inst{3} Institute of Applied Physics, Xi'an University of Arts and Science, Xi'an 710065, China\\
  \inst{4} School of Information and Communication Engineering, Tianjin Polytechnic University, Tianjin 300160, China \\
  \inst{(a)} Corresponding author: zhangyiqi@mail.xjtu.edu.cn\\
  \inst{(b)} Corresponding author: ypzhang@mail.xjtu.edu.cn
}

\abstract{
  We demonstrate that beams originating from Fresnel diffraction patterns are self-accelerating in free space.
  In addition to accelerating and self-healing, they also exhibit parabolic deceleration property,
  which is in stark difference to other accelerating beams.
  We find that the trajectory of Fresnel paraxial accelerating beams is similar to that of nonparaxial Weber beams.
  Decelerating and accelerating regions are separated by a critical propagation distance,
  at which no acceleration is present. During deceleration,
  the Fresnel diffraction beams undergo self-smoothing, in which
  oscillations of the diffracted waves gradually smooth out and are completely gone at the critical distance.
  }

\begin{document}

\maketitle

Accelerating beams in free space or in linear dielectric media have attracted a lot of attention in the past decade, owing to their
interesting properties which include self-acceleration, self-healing, and non-diffraction over many
Rayleigh lengths \cite{siviloglou_ol_2007,siviloglou_prl_2007,broky_oe_2008, ellenbogen_np_2009,chong_np_2010}.
Because Airy function is the solution of the linear Schr\"odinger equation \cite{berry_ajp_1979,lin_epl_2008},
the reported paraxial accelerating beams are all related to the Airy or Bessel functions \cite{durnin_josaa_1987,bouchal_cjp_2003}.
Nonparaxial accelerating beams, for example Mathieu and Weber waves, are found by solving Helmholtz wave equation \cite{zhang_prl_2012}.
To display acceleration such beams must possess energy distributions which lack parity symmetry in the transverse direction.

Airy accelerating beams have also been discovered in nonlinear media \cite{kaminer_prl_2011,lotti_pra_2011,panagiotopoulos_pra_2012},
Bose-Einstein condensates \cite{efremidis_pra_2013},
on the surface of a gold metal film \cite{minovich_prl_2011} or on the surface of silver \cite{li_prl_2011,li_nano_2011}, in
atomic vapors with electromagnetically induced transparency \cite{zhuang_ol_2012b},
chiral media \cite{zhuang_ol_2012a}, photonic crystals \cite{kaminer_oe_2013}, and elsewhere.
A wide range of applications of accelerating beams has already been demonstrated, for example, for
tweezing \cite{garces_nature_2002}, the generation of plasma channels \cite{fan_prl_2000}, material modifications \cite{amako_josab_2003},
microlithography \cite{erdelyi_jvs_1997}, light bullet production \cite{chong_np_2010},
particle clearing \cite{baumgartl_np_2008}, and manipulation of dielectric microparticles \cite{zhang_ol_2011}.

In this Letter we search for a new kind of accelerating beams -- those generated in the paraxial propagation of Fresnel diffraction patterns.
We display self-accelerating and self-healing properties of such beams, first in one dimension (diffraction from a straight edge),
and then in two dimensions (diffraction from a corner).
We demonstrate that there exists a critical propagation distance, at which acceleration stops;
before the critical distance the diffraction patterns decelerate and after the critical distance they accelerate.
During the deceleration phase the interference oscillations are suppressed, owing to the self-smoothing effect.
Both the deceleration and the acceleration phases exhibit parabolic trajectories,
which appear to be similar to the trajectories of nonparaxial Weber accelerating beams.

\begin{figure}[htbp]
  \centering
  \includegraphics[width=\columnwidth]{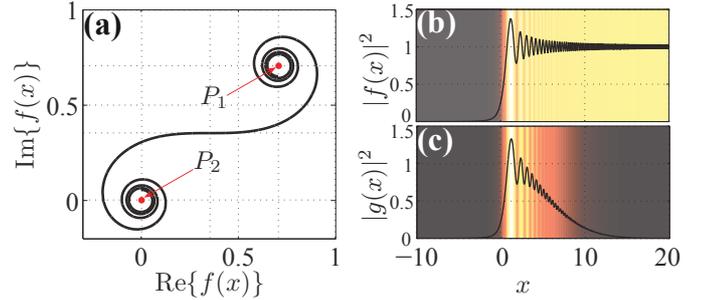}
  \caption{(Color online) (a) Cornu spiral of $f(x)$ in Eq. (\ref{psi1}).
  (b) Intensity $|f(x)|^2$ versus $x$.
  (c) Intensity $|g(x)|^2$ versus $x$, with $a=0.01$.
  The background images in (b) and (c) depict the ideal and attenuated diffraction stripes.}
  \label{fig1}
\end{figure}

We begin our analysis by briefly recalling Fresnel diffraction of plane waves from a straight edge located at $x=0$,
which can be viewed as a one-dimensional (1D) case. We assume $x$ is the transverse coordinate and $z$ the propagation direction.
The normalized amplitude of the diffraction pattern is described by \cite{born_book}
\begin{align}\label{psi1}
  f(x)=\frac{1}{\sqrt{2}} \left[\left(\mathcal{C}(x)+\frac{1}{2}\right)+i\left(\mathcal{S}(x)+\frac{1}{2}\right)\right],
\end{align}
in which
\begin{align}
\mathcal{C}(x)=&\int^x_0 \cos\left(\frac{\pi}{2}\tau^2\right)d\tau, ~\rm and \notag\\
\mathcal{S}(x)=&\int^x_0 \sin\left(\frac{\pi}{2}\tau^2\right)d\tau \notag
\end{align}
are the Fresnel Cosine and Sine integrals \cite{abramowitz_book}. The limiting values of the Sine and Cosine
integrals are $\mathcal{C}(\infty)=\mathcal{S}(\infty)=\frac{1}{2}$ and $\mathcal{C}(-\infty)=\mathcal{S}(-\infty)=-\frac{1}{2}$, and
the real and imaginary parts of $f(x)$ in Eq. (\ref{psi1}) form the Cornu spiral, as shown in Fig. \ref{fig1}(a).
The two branches of the spiral approach the points $P_1$ and $P_2$ with coordinates $(\frac{1}{\sqrt{2}},\frac{1}{\sqrt{2}})$ and $(0,0)$, respectively.
Since the ideal $f(x)$ is not square-integrable, that is $\int_{-\infty}^{+\infty} |f(x)|^2 dx \rightarrow \infty$,
it possesses infinite energy, which is not very realistic. However, that's not unusual.

The initial appearance of accelerated beams \cite{berry_ajp_1979,durnin_josaa_1987}
attracted some controversy, because they were nondiffracting but also
of infinite energy in free space and as such not much
physically realistic. However, the same features are shared by
plane waves, which are also unrealistic yet very useful.
The necessity of having
finite apertures and finite-size lenses in the production of nondiffracting beams meant
that some diffraction must be present. By now, this initial controversy
has settled and nondiffracting Airy and Bessel
optical beams have become vibrant part of linear optics. However,
no such controversy exists in nonlinear optics, where nondiffracting
localized beams -- solitons -- commonly appear.
Hence, we introduce a Gaussian aperture function $\exp(-ax^2)$, to make $f(x)$ square-integrable;
the modified $f(x)$ is written as
\begin{align}\label{psi2}
  g(x)=\frac{1}{\sqrt{2}} \left[\left(\mathcal{C}(x)+\frac{1}{2}\right)+i\left(\mathcal{S}(x)+\frac{1}{2}\right)\right] \exp(-ax^2),
\end{align}
in which $a>0$ is the decay factor that connects with the numerical aperture of the system.

In Figs. \ref{fig1}(b) and \ref{fig1}(c), we display the intensity profiles
as well as numerically observed interference stripes present in $f(x)$ and $g(x)$.
It is seen that the oscillating tail of $f(x)$ tends to 1 as $x\rightarrow \infty$,
whereas the tail of $g(x)$ tends to 0, which assures finite energy of the wave packet.
In addition, the energy distribution of $g(x)$ is asymmetric, which assures self-acceleration of the wave packet
\cite{siviloglou_ol_2007}.

The linear Schr\"odinger equation for the slowly-varying envelope of the paraxial wavepacket in free space
or in linear dielectric media in 1D can be written as
\begin{equation}\label{eq3}
  \frac{\partial g}{\partial z} + \frac{1}{2} \frac{\partial^2 g}{\partial x^2}=0,
\end{equation}
in which $x$ and $z$ coordinates are normalized to some typical transverse size of localized beams $x_0$ and to
the corresponding Rayleigh range $k x_0^2$, respectively.
One of the solutions to this equation is the celebrated Airy beam \cite{berry_ajp_1979}
\begin{align*}
  g(x,z)={\rm Ai}\left(x-\frac{z^2}{4}\right)\exp\left(\frac{6xz-z^3}{12}\right),
\end{align*}
which opened the whole field of linear nondiffracting beams. For difference, we input Fresnel finite diffraction
pattern $g(x)$ into Eq. (\ref{eq3}) and consider what happens.

\begin{figure}[htbp]
  \centering
  \includegraphics[width=\columnwidth]{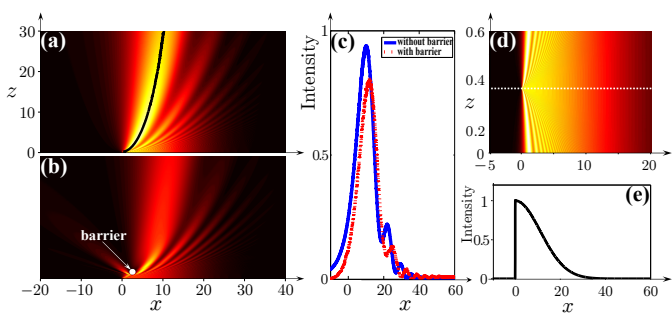}
  \caption{(Color online) (a) Propagation of $g(x)$. The solid curve depicts the acceleration of the main lobe.
  (b) Self-healing of the beam, visible when the main lobe of $g(x)$ is blocked by a circular barrier (the white dot).
  (c) Solid and dashed curves are intensities at $z=30$ corresponding to (a) and (b), respectively.
  (d) Same as (a), but for a much shorter propagation distance. The deceleration phase of the propagation is clearly visible.
  (e) Intensity profile at the critical propagation distance, marked by the dashed line in (d).
  The value of $a$ is 0.002 in all the cases shown.}
  \label{fig2}
\end{figure}

The evolution of the finite-energy diffraction pattern is shown in Fig. \ref{fig2}(a).
It is seen that the beam accelerates to the right, even though it propagates in the linear medium.
In fact, the intensity maximum of the beam accelerates during propagation along a parabolic trajectory, as shown by the solid curve.
This is characteristic of all self-accelerating linear beams: high-intensity portions of the beam accelerate, while the center of mass
of the beam actually moves along a straight line \cite{bouchal_cjp_2003}.
But, different from the previous observations \cite{siviloglou_ol_2007,siviloglou_ol_2008,siviloglou_prl_2007},
in which the Airy beam accelerates according to the asymptotics $x \propto z^2$,
the diffraction pattern accelerates according to $x^2 \propto z$.
This is similar to the accelerating trajectory of a nonparaxial Weber beam \cite{zhang_prl_2012,bandres_njp_2013}.

The self-healing can be seen clearly if a small barrier of size $0.5$ is placed in the path of the main lobe propagation, at $z=0$;
the corresponding evolution is shown in Fig. \ref{fig2}(b).
The output intensity profiles with and without the barrier, shown in Fig. \ref{fig2}(c),
illustrate that the self-healing of the main lobe is apparent.

It is worth noting that
there appears a new phase at a short propagation distance.
To see the phenomenon clearly, the short distance propagation from Fig. \ref{fig2}(a) is enlarged in Fig. \ref{fig2}(d).
We find that the diffraction pattern undergoes $-x^2 \propto z$ acceleration firstly,
before accelerating according to $x^2 \propto z$ afterwards.
Therefore, the initial acceleration process is actually a deceleration.
During the deceleration process, the oscillations are suppressed gradually,
which actually represents a self-smoothing effect.
It can be explained phenomenologically.

During deceleration, the interference fringes accumulate at $x=0$ point.
Since diffraction stripes cannot appear in the $x<0$ region,
$x=0$ will be a decelerating destination for all the lobes. At that point along the $z$ axis -- the critical
point -- the transverse motion stops, fringes are gone, and
the beam profile becomes smooth. This smooth intensity profile is shown in Fig. \ref{fig2}(e),
recorded at the critical distance, marked by the dashed line in Fig. \ref{fig2}(d).
The profile looks like a 1D Gaussian beam truncated by the Heaviside step-function.
After the critical propagation distance,
the oscillations reappear and display the usual $x^2 \propto z$ acceleration.
If one looks at the motion of the center of mass of the beam, during deceleration
it approaches steadily the $x=0$ wall, bounces off, and continues to move steadily away,
as the beam accelerates.

\begin{figure}[htbp]
  \centering
  \includegraphics[width=\columnwidth]{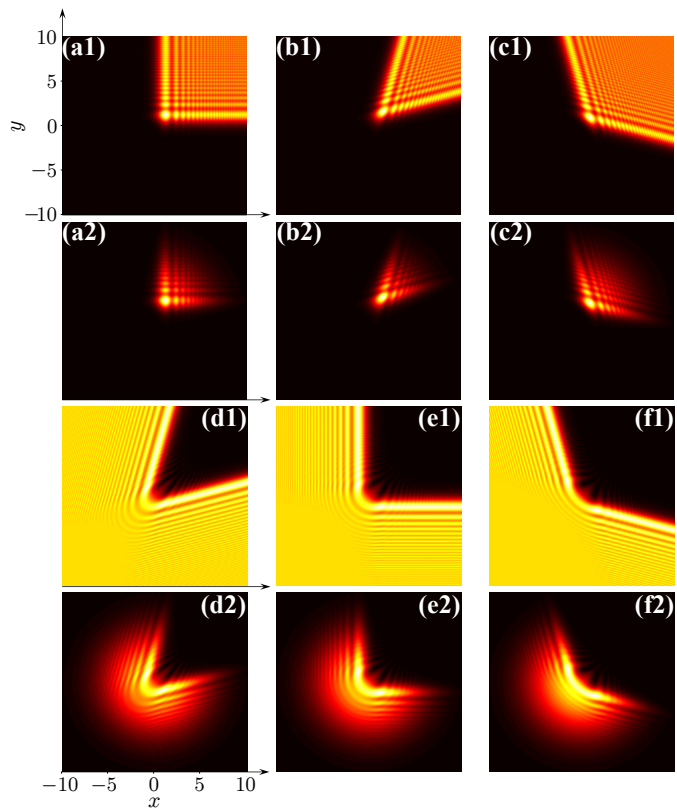}
  \caption{(Color online) (a1) Fresnel diffraction pattern at a right-angle corner.
  (b1) and (c1) Fresnel diffraction patterns with $\theta=\pi/3$ and $2\pi/3$, respectively.
  (a2)-(c2) Truncated Fresnel diffraction patterns according to (a1)-(c1) with $a=0.02$, respectively.
  (d1)-(f1) Fresnel diffraction patterns from a wedge with angles $\pi/3$, $\pi/2$, and $2\pi/3$, respectively.
  (d2)-(f2) Truncated diffraction patterns corresponding to (d1)-(f1).}
  \label{fig3}
\end{figure}

We now analyze the accelerating properties of a two-dimensional (2D) diffraction pattern,
by propagating Fresnel diffraction from a right-angle corner located at $(x=0,y=0)$.
The corresponding diffraction pattern is described by
\begin{align}\label{corner}
  F(x,y) = & \frac{1}{2} \left[\left(\mathcal{C}(x)+\frac{1}{2}\right)+i\left(\mathcal{S}(x)+\frac{1}{2}\right)\right] \times \notag \\
           & \left[\left(\mathcal{C}(y)+\frac{1}{2}\right)+i\left(\mathcal{S}(y)+\frac{1}{2}\right)\right],
\end{align}
according to Eq. (\ref{psi1}).
To make the wave packet of finite energy, we still introduce a Gaussian decaying aperture,
so that Eq. (\ref{corner}) is modified as
\begin{align}\label{corner2}
  G(x,y) = F(x,y) \exp\left[-a(x^2+y^2)\right].
\end{align}
The diffraction patterns based on Eqs. (\ref{corner}) and (\ref{corner2}) are shown in Figs. \ref{fig3}(a1) and \ref{fig3}(a2), respectively.
It is clear that the ideal 2D diffraction pattern is not square integrable,
whereas the truncated one is finite-energy, similar to the 2D finite-energy Airy beam \cite{siviloglou_ol_2007}.
Furthermore, the right-angle corner diffraction can be easily generalized to
2D acute or obtuse angle Fresnel diffraction, as shown in Figs. \ref{fig3}(b) and \ref{fig3}(c).
This could be done, for example, by utilizing the Lorentz transformation of coordinates \cite{eichelkraut_ol_2010}:
\begin{align*}
  x'= & x\cosh[-(1/2)\tanh^{-1}(\cos\theta)]+ \\
      & y\sinh[-(1/2)\tanh^{-1}(\cos\theta)], \\
  y'= & x\sinh[-(1/2)\tanh^{-1}(\cos\theta)]+ \\
      & y\cosh[-(1/2)\tanh^{-1}(\cos\theta)],
\end{align*}
where $0<\theta<\pi$ is the angle at the corner, and $x'$ and $y'$ the oblique axes coordinates.
Substituting $(x,y)$ by $(x',y')$ in Eqs. (\ref{corner}) and (\ref{corner2}),
one transforms the Fresnel integrals into the oblique diffraction patterns.
Figures \ref{fig3}(b1) and \ref{fig3}(c1) represent ideal diffraction patterns,
while Figs. \ref{fig3}(b2) and \ref{fig3}(c2) show the corresponding truncated ones.

If the angle of the corner is bigger than $\pi$,
it can be viewed as a diffraction from a corner of a wedge.
For this case, the analytical expression for an ideal diffraction pattern can be written as
\begin{align}\label{disk}
  F(x,y) = & \frac{1}{2} \left[\left(-\mathcal{C}(x)+\frac{1}{2}\right)+i\left(-\mathcal{S}(x)+\frac{1}{2}\right)\right] \times \notag \\
           & \left[\left(\mathcal{C}(y)+\frac{1}{2}\right)+i\left(\mathcal{S}(y)+\frac{1}{2}\right)\right] + \notag \\
           & \frac{1}{2} \left[\left(\mathcal{C}(x)+\frac{1}{2}\right)+i\left(\mathcal{S}(x)+\frac{1}{2}\right)\right] \times \notag \\
           & \left[\left(-\mathcal{C}(y)+\frac{1}{2}\right)+i\left(-\mathcal{S}(y)+\frac{1}{2}\right)\right] + \notag \\
           & \frac{1}{2} \left[\left(-\mathcal{C}(x)+\frac{1}{2}\right)+i\left(-\mathcal{S}(x)+\frac{1}{2}\right)\right] \times \notag \\
           & \left[\left(-\mathcal{C}(y)+\frac{1}{2}\right)+i\left(-\mathcal{S}(y)+\frac{1}{2}\right)\right].
\end{align}
Based on this formula and the Lorentz transformation,
one can obtain the 2D Fresnel diffraction pattern from a wedge with an angle $0<\theta<\pi$.
In Figs. \ref{fig3}(d)-\ref{fig3}(f),
we display diffraction patterns with $\theta$ being $\pi/3$, $\pi/2$, and $2\pi/3$, respectively.

\begin{figure}[htbp]
  \centering
  \includegraphics[width=\columnwidth]{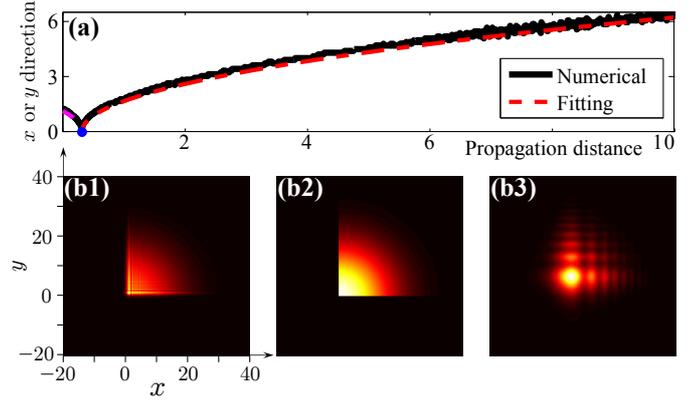}
  \caption{(Color online) (a) Decelerating and accelerating trajectories of the two-dimensional truncated diffraction pattern for $a=0.002$.
  The dot ($\color{blue}{\bullet}$) marks the critical distance.
  Solid and dashed curves correspond to the numerical and fitted results.
  (b1)-(b3) Two-dimensional finite-energy diffraction beams shown at the input, at the critical point, and at the output, respectively.}
  \label{fig4}
\end{figure}

For the 2D propagation case, Eq. (\ref{eq3}) should be modified into
\begin{equation}\label{eq5}
  \frac{\partial G}{\partial z} + \frac{1}{2} \left( \frac{\partial^2}{\partial x^2} +  \frac{\partial^2}{\partial y^2}\right) G=0.
\end{equation}
By inputting the diffraction pattern from Fig. \ref{fig3}(a2) into Eq. (\ref{eq5}),
the evolution of the truncated 2D Fresnel diffraction pattern can be investigated.
We consider right away the more interesting case where a small circular barrier is placed diagonally, to block the propagation
of the main lobe. In Fig. \ref{fig4}(a),
we exhibit the evolution trajectory of the main lobe that is projected onto $x0z$ or $y0z$ plane by the solid curve.
It is seen that the pattern displays acceleration along a parabolic profile
and that there is still a critical propagation distance, marked by the dot ($\color{blue}{\bullet}$) in Fig. \ref{fig4}(a).
To the left and right of the dot, two pieces of parabola are seen.
To roughly fit the numerically obtained decelerating and accelerating trajectories, we introduce two ansatzes as
\begin{align*}
  x_1= & 2\sqrt{{\color{blue}{\bullet}}-z_1}, \\
  x_2= & 2\sqrt{z_2-\color{blue}{\bullet}},
\end{align*}
which are depicted in the figure by the dashed curves. As is evident,
the ansatzes fit the numerical results quite well.
We note that the fluctuations seen in the solid curve result not from the oscillations in the profiles
but from not having high enough numerical accuracy.
These fluctuations do not affect the decelerating or accelerating trends visible overall.

In Figs. \ref{fig4}(b1)-\ref{fig4}(b3),
we show the intensity images of the 2D diffraction pattern at the
input ($z=0$), the critical propagation distance, and the output ($z=10$), respectively.
Since there are no oscillations in the beam at the critical distance,
the self-smoothing effect is still in effect.
The maximum intensity profile, located at $(x=0,y=0)$,
which is the decelerating destination of all the lobes, is still described by
the 1D case shown in Fig. \ref{fig2}(e). In addition,
similar results hold for the cases shown in Figs. \ref{fig3}(d2)-\ref{fig3}(f2);
they also display a critical propagation distance and the self-smoothing effect.
Therefore, we do not discuss here the corresponding numerical results.

In conclusion,
we have demonstrated that Fresnel diffraction patterns can be viewed as accelerating beams,
which also exhibit self-accelerating and self-healing properties.
Different from Airy accelerating beams, the new accelerating beams introduced in this Letter
exhibit deceleration and strong self-smoothing effect at the critical propagation distance.
Right at the critical distance the oscillations in the Fresnel diffraction beam disappear,
and beyond this point the oscillations reappear again.
It is worth noticing that the accelerating process follows parabolic trajectory, similar to
Weber beams; however, the propagation can be divided into two regions.
Before the critical propagation distance, the beam decelerates according to $-x^2 \propto z$;
after the critical point, the beam accelerates according to $x^2 \propto z$.
Our research not only demonstrates
a new kind of accelerating beam, but also broadens people's
understanding on Fresnel diffraction.

\acknowledgments
This work was supported by CPSF (2012M521773),
the Qatar National Research Fund NPRP 09-462-1-074 project,
the 973 Program (2012CB921804), NSFC (61078002, 61078020, 11104214, 61108017, 11104216, 61205112),
RFDP (20110201110006, 20110201120005, 20100201120031),
and FRFCU (xjj2013089, 2012jdhz05, 2011jdhz07, xjj2011083, xjj2011084, xjj2012080).


\end{document}